\documentstyle[epsf]{elsart}
\input{psfig.sty}
\newfont{\cmsy}{cmsy10}
\begin{document}
\begin{frontmatter}

\vspace*{-2.0cm}

\bigskip

\begin{center}
{\Large {\bf 
	SEARCH FOR DCC IN RELATIVISTIC  \\
        HEAVY-ION COLLISION THROUGH EVENT \\
        SHAPE ANALYSIS\\
}}
\vspace*{0.2cm}
{B. K. Nandi, G. C. Mishra, B. Mohanty, 
and D. P. Mahapatra}\\
Institute of Physics, Bhubaneswar, India\\
\vspace*{0.1cm}
{T. K. Nayak} \\
Variable Energy Cyclotron Centre, Calcutta, India \\ 
\end{center}

\normalsize

\begin{abstract}
	Event shape analysis has been used to look for DCC signals
	in simulated ultra-relativistic heavy-ion collision data at 
	SPS energy. A simple redistribution of particles, with two 
	detectors to detect charged particles and photons, is seen
	to result in the same flow direction with the flow angle
	difference peaking at zero. However, events where the neutral 
	pion fraction has been modified according to the DCC probability 
	distribution, show the flow angles in two detectors to be almost
	$90^o$ apart. The results presented here show that the technique
	is complementary to the one based on the discrete wavelet
	transformation. Together the techniques are seen to provide
	a very powerful tool for DCC search in ultra relativistic
	heavy ion collision.

\end{abstract}
\end{frontmatter} 
\section{Introduction}

	In heavy-ion collisions at ultra-relativistic energies there
is a rapid expansion of the collision debris in the longitudinal direction
leading to a super cooling
of the interior interaction region. This is expected to lead to the
formation of unconventionally oriented vacuum structures 
as allowed by the chiral symmetry \cite{bjor1,blai1,raj1}. 
These are called the Disoriented Chiral Condensates (DCC).
DCC formation results in large fluctuation in the neutral pion fraction.
The probability distribution of the neutral pion fraction is 
characterised by,
\begin{equation}
P(f)={1\over {2\sqrt{f}}}
\end{equation}

\begin{equation}
{\rm where~~~~} f={N_{\pi^0}\over {N_{\pi^0}+N_{\pi^+}+N_{\pi^-}}}
\sim {{N_{\gamma}/2}\over {N_{\gamma}/2+N_{ch}}}
\end{equation}

$N_\gamma$ and $N_{\rm ch}$ are multiplicities for photons and charged particles,
respectively. This assumes all charged particles are pions and all photons come from
$\pi^0$ decay.

Detection of this interesting phenomenon of DCC formation is expected
to provide valuable information on the vacuum structure of strong
interaction and chiral phase transition. Therefore, it has become a
very interesting aspect of heavy ion collision studies and a number of
experiments have been planned at RHIC and LHC energies. 

In an experiment involving the search of localised DCC one essentially 
tries to detect a state with a large and 
localised fluctuation in the ratio of the number of
photons to charged particles. In principle 
there should be regions in the detected phase space where the number $f$
should be very much different from 1/3. A typical event structure would
be very similar to the anti-centauro event as reported by the JACEE
collaboration \cite{jacee}. In view of this, a typical experiment 
would consist of two detectors to detect the charged particles and the 
photons respectively. These two detectors should have complete
overlap in $\eta$ and $\phi$ with as much $\eta$ coverage as possible. 
From the detected hit patterns of charged and neutral particles one tries
to see whether there is any local fluctuation of $f$ indicating
the signature of localised DCC.

	As far as DCC search goes many analysis methods have been
proposed \cite{bjor1,wa98dcc,huang,nandi,tapan1,brooks}. However, the most attractive 
method has been the one based
on the Discrete wavelet transforms (DWT) proposed by Huang et al.
\cite{huang}. This has the beauty of
analysing a spectrum at different length scales with the ability of
finally picking up the right scale at which there is a fluctuation.
Because of this advantage, although there has been no claim regarding 
the observation of DCC, the method was successfully applied to filter
out interesting events with large photon to charge particle fluctuation
in WA98 experiment \cite{nandi,tapan1}. 

In the present paper, using simulated data, attempts have been made to 
show the power of yet another analysis technique which has been used 
successfully for flow analysis in heavy-ion data \cite{wa93flow,olitr}. 
The method is based on the simple fact that localised DCC formation 
is expected to lead to an event shape anisotropy which is expected 
to be out of phase for one detector corresponding to the other. 
In other words it means, whenever there is large number of charge 
particles recorded in a DCC region in one detector, there should be 
a depletion in the number of neutral particles in the same zone of 
the other detector. Therefore, from an event shape analysis using 
both detectors one can, in principle, look for DCC signals. 
On the other hand, in the DWT analysis, one tries to look at the 
neutral pion fraction 
$f$ at different resolutions constructing what are known as the Father 
function coefficients (FFC) and Mother function coefficients (MFC). 
Without DCC, the FFC distribution for simulated generic events has been
shown to be a Gaussian. However with DCC, the distribution goes to a
non-Gaussian shape with several events appearing in the wings \cite{nandi}
The events which lie beyond the generic Gaussian can be picked up 
as DCC-like events. One can also construct the power spectrum for the
FFCs and look for their variation as a function of the length scale.
For generic events with only a statistical fluctuation in the numbers
of charged and neutral pions the FFC power spectrum shows a flat 
curve without a structure at any scale. However, with DCC-like
fluctuation, generated over a given domain of phase space, the power
spectrum is expected to show an enhancement at scales below the
specified domain size. 

In the present case we have carried out simulation of DCC-like and
pure flow-type events and have applied the technique of event shape
analysis which distinguishes very clearly both class of events. In
case of simulated DCC-like events, particularly when their fraction
in a large number of events is comparatively small, the present method, 
together with that based on DWT, has been found to be successful in
finding out DCC-type signature. 

\section{Modeling DCC and event anisotropy}

For DCC production, a procedure which is similar to the 
one employed in \cite{wa98dcc} has been followed. VENUS 4.12 event 
generator \cite{venus} has been used for the simulation of DCC type events at SPS 
energy ($Pb$ on $Pb$). In this, the charge of the pions is interchanged 
pairwise ($\pi^{+}\pi^{-} \leftrightarrow \pi^{0}\pi^{0}$), 
in a selected $\eta-\phi$ zone according to the DCC probability 
distribution as given in equation (1) event by event. Finally the 
$\pi^0$s are allowed to decay. 

	For the present study, DCC events have been simulated in a
range $3.0\le \eta \le 4.0$ with a domain size having 
$\Delta \phi = 90^o$. For the analysis 20,000 events were generated.
A similar amount of VENUS events were also generated for
comparison. However, to simulate what happens in a true experimental
situation it is also necessary to include detector related effects.
For photon and charge
particle detection the detection efficiencies were taken to be about
$70\pm 5 \%$ and $95 \pm 2 \%$ respectively. It is also known that
charged particles sometimes lead to photon-like signals and such a 
contamination in an experimental situation can be as high as 25 \% 
\cite{wa98nim}. 
Following this, it was decided to include a 25 \% charged particle
contamination in the photon signal. Finally, we have
also prepared several sets of data with different DCC fractions 
(10 $-$ 100 \%) mixing generic VENUS and DCC type events in an appropriate 
manner.

To introduce event anisotropy in every event, a simple toy model
has been employed. Here the distributions for both charged and 
neutral particles are generated from VENUS,
distorted according to a procedure as given below. First of all 
a flow direction is selected at random, distributed uniformly between 
$0^o$ and $360^o$. 
About each flow direction, corresponding to a given 
event, a Gaussian particle distribution, with a $\sigma$ of 
$10^o$ is generated by picking VENUS 
generated particles at random. Here the charge conservation in every 
localised region is ensured since all three types of particles are
selected with equal probability.

	Constructed as above, both localised DCC and simple event
anisotropy (indicating flow) are modeled in different events and the
results were analysed using the standard flow analysis as employed
elsewhere \cite{olitr}. In the present study the method of Fourier 
analysis with n=2 (elliptic flow) has been employed. 

\section{Method of analysis}

The particle distribution in a given detector can be written as a set 
points ($\eta_{i},\phi_{i},i=1,N$) showing the hit pattern in the 
detector. 

\noindent In the second order elliptic flow, for each event, one tries to 
construct the sums

\begin{equation}
      X = \Sigma cos(2\phi_i)
\end{equation}
\begin{equation}
	Y= \Sigma sin(2\phi_i)
\end{equation}

\noindent The flow angle $\Phi$ is determined from these two sums using the
expression 

\begin{equation}
	\Phi = {1\over 2}tan^{-1}(Y/X)
\end{equation}

In the absence of any detector imperfections and other geometrical
effects, the distribution of $\Phi$, taken over a large number of events
is expected to be flat without any peaks or bumps spread over $0^o$ to
$180^o$. This is because the flow direction varies randomly from
event to event. However, when the events are realigned, with respect to
the flow angle in each event, one can see the characteristic peaks
(at $0$ and $180^o$) in the azimuthal distribution of particles.

In case of two detectors with the same phase space ($\eta-\phi$) coverage,
one detecting photons and other detecting charged particles 
the situation is very interesting. If there is 
genuine flow in a particular event both detectors would show the effect 
in terms of $\Phi$ angles getting aligned in the same direction. Therefore the 
distribution of $\Psi(=\Phi_1- \Phi_2)$, the difference between the flow 
angles, as obtained for the two detectors is expected to be peaking at zero.
However, in case of DCC being prominent in a particular region, there
will be more photons detected in one detector. The other detector is 
expected to show less charged particles in the same region of phase space.
Therefore an event shape analysis is expected to show two flow angles
for both detectors which will be out of phase (with n=2 the angular
difference is expected to peak at $90^o$). This is the most important
result based on which the present analysis is carried out.

\section{Results and Discussions}
Results shown in Fig.1 (a-c) correspond to the case with only DCC
in which one can notice (Fig. 1 a) a clear peak
for the angle between the event planes for the two detectors, $\Psi$,
at $90^o$. Figs. 1 b and c show the angle between the event planes
as obtained for two sub-events in each of the detectors separately
It is important to notice that both detectors show "flow", but one
with respect to the other clearly shows an anti-flow type behavior.
In the same figure we have also presented the data corresponding
to pure VENUS events for comparison. One can notice there is neither
any signature of flow nor DCC-like fluctuation.

Results shown in Figs. 1 (d-f) show the same plots for flow type
events which have no DCC-like fluctuation. Here, the individual 
detectors are seen to show the same effect as one with respect to the
other.

In Fig.2 we have presented event shape analysis results for 20000
events having different fractions of DCC-type fluctuation ranging 
from 10 \% to 100 \%. One can notice that the expected peak around 
$90^{o}$ gets weaker as DCC fraction decreases. This is primarily 
because of an overwhelming majority of generic events that contribute
uniformly to the $\Psi$ distribution over the entire angular range.
So in order to look for any DCC-like signature one has to suppress the 
contribution of these events filtering out the interesting DCC-like
ones. 

In Fig.3 we have shown the FFC distribution for pure generic
and pure DCC-like events at scale $j = 1$. We find that the FFC 
distribution for DCC-like events is broader in comparison to
that for generic events. In fact, one can show that there is pile up of
a large number of events within the width of the generic distribution
 with decrease in the fraction of DCC-like events. But it is
those events that lie beyond the width of the generic distribution, 
about which some definite conclusion regarding their DCC-like nature 
can be made. In such a case, when a fraction of the events contain
the DCC type signal, it is very difficult to notice their signature
using the power spectrum analysis \cite{nandi} which employs the 
event averaging procedure. This is because the interesting events 
lying beyond the width of the generic distribution, contributing
significantly to the power spectrum, are overwhelmingly outnumbered 
by those lying within the width.

A simple method to filter out the contribution of a great majority of 
non-DCC type events is to apply a cut on the width of the FFC distribution 
and consider the events lying above the cut. With this in view, in the 
present case, we have applied a cut of $\pm 1.5 \sigma$ in the FFC 
distribution. The $\Psi$ Distribution of the filtered events for the 
case of 25 \% of DCC-like events is shown in Fig.4. One can clearly 
notice the peaking at $\Psi = 90^{o}$
indicating a DCC-like signature. The pure VENUS events with the same cut in their
FFC distribution are also shown for comparison. They are seen to contribute
uniformly over the entire angular range.

This clearly shows that the DWT
and the event shape analysis applied together can be a very powerful
method for the search of DCC. However, one needs to judiciously use an 
appropriate cut on the FFC distribution to filter out a great many 
uninteresting effects. Before concluding, a word must be mentioned 
regarding the errors.  One can notice, the spectral distribution 
shows a histogram where the error in the entry at every angle goes 
as $\sqrt{N}$. By filtering out a large number of events, which are 
mostly distributed uniformly over the entire angular range, one essentially 
reduces the background although the statistical errors increase slightly.

\section {Conclusion}

In the present paper it has been demonstrated that the technique 
of event shape analysis can be very effectively employed to look for 
the signature of DCC formation in relativistic heavy ion collision 
data. Here, in the absence of any rigorous theoretical model to 
simulate DCC formation, an isospin fluctuation has been introduced
locally on the VENUS generated events to generate a charged to neutral 
pion asymmetry. At least for the case with a single, large, DCC domain,
it seems to be a very effective technique for looking at DCC signature.
However, in cases where only a certain fraction of the events are
expected to be of DCC type, together with the technique of DWT which
provides a first hand filter, the present technique provides a very
powerful probe for DCC signal.

\newpage
\noindent
{\Large {\bf
        Figure Captions  \\
}}
\vspace*{0.2cm}
\normalsize

\noindent
{\bf Fig.1:} The distribution of $\Psi$ for simulated DCC(a-c) and
             flow(d-f). (a) shows anti-correlation between
             event planes determined for charged particle and photon
             detectors. (b) and (c) show correlations between event
             planes obtained considering two sub events for the photon
             and charged particle detectors respectively.
             The dotted lines in figures (a-c) show the distributions
             for generic VENUS events. In case of flowy events
             (d) shows correlation between event planes obtained from
             the two detectors. (e) and (f) are same as (b) and (c). \\

\noindent
{\bf Fig.2:} The distribution of $\Psi$ for simulated DCC and VENUS
	     with varying DCC fractions 
	     (a) 100 \% of the events DCC
	     (b) 75 \% of the events DCC 
	     (c) 50 \% of the events DCC
             (d) 25 \% of the events DCC  
	     (e) 15 \% of the events DCC
             (f) 10 \% of the events DCC   \\

\noindent
{\bf Fig.3:} FFC distribution for simulated generic VENUS and pure
             DCC-like events. The solid and the dotted lines correspond
             to the DCC-like and VENUS events respectively.\\

\noindent

{\bf Fig.4:} The distribution of $\Psi$ as obtained
             from the two detectors excluding events lying within 1.5
	     sigma of the FFC distribution of VENUS. The dotted lines
	     shows the distribution of $\Psi$ as above but for pure VENUS
	     events. Errors though not
	     shown are of the order of $\sqrt{N}$ at every point. \\

\newpage


\noindent ${}$\hbox{\epsffile{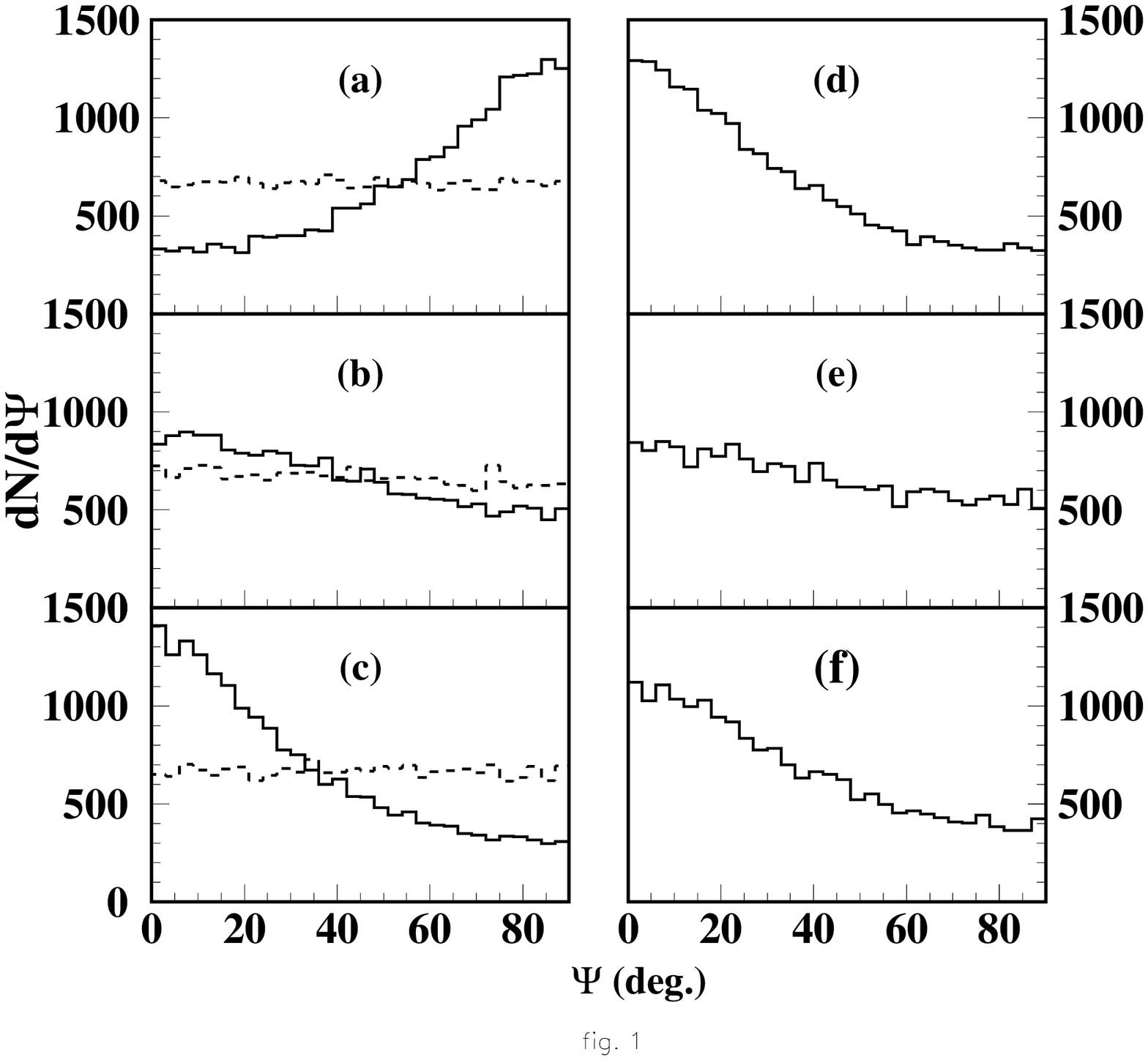}}
\newpage

\noindent ${}$\hbox{\epsffile{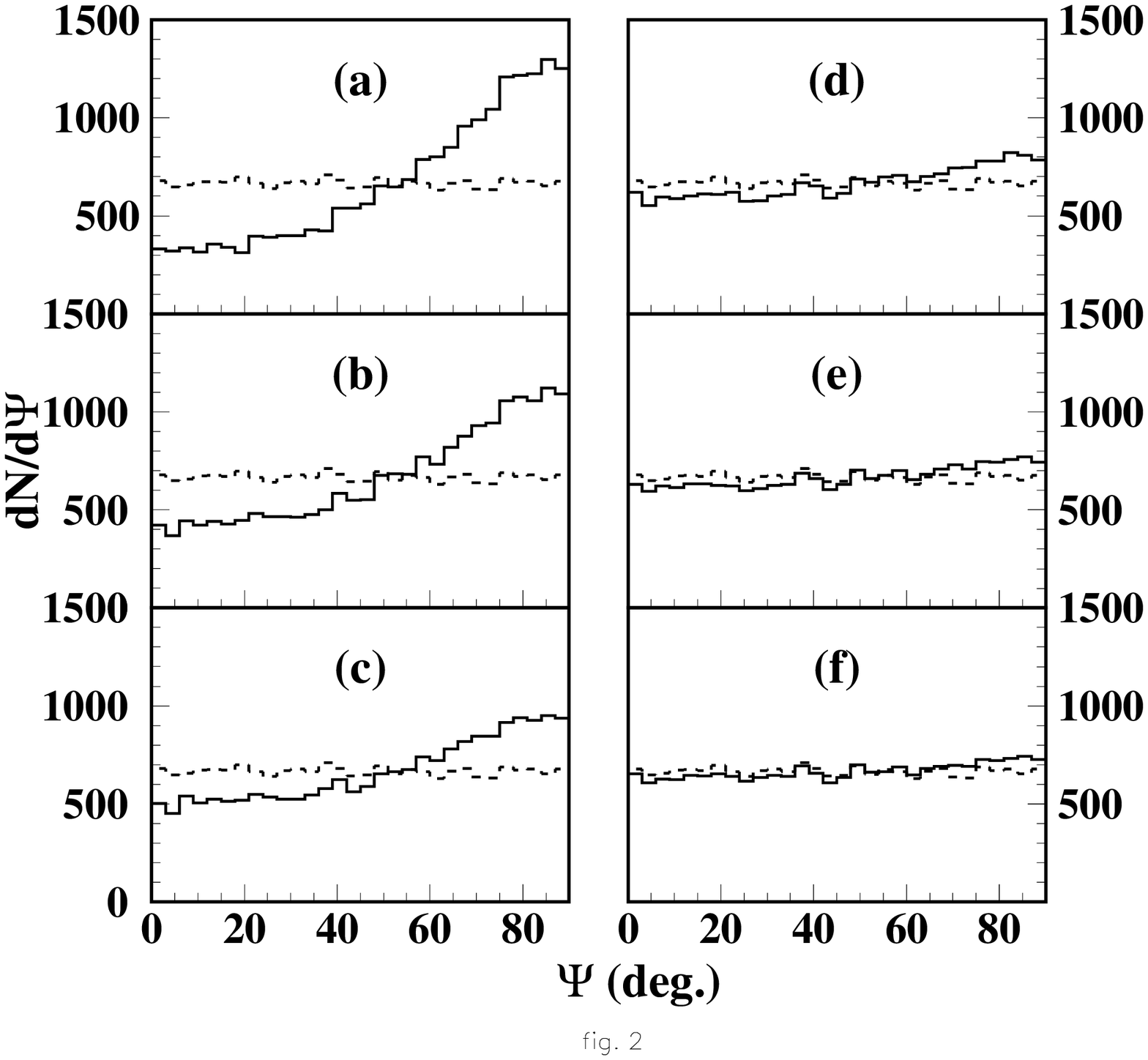}}
\newpage

\noindent ${}$\hbox{\epsffile{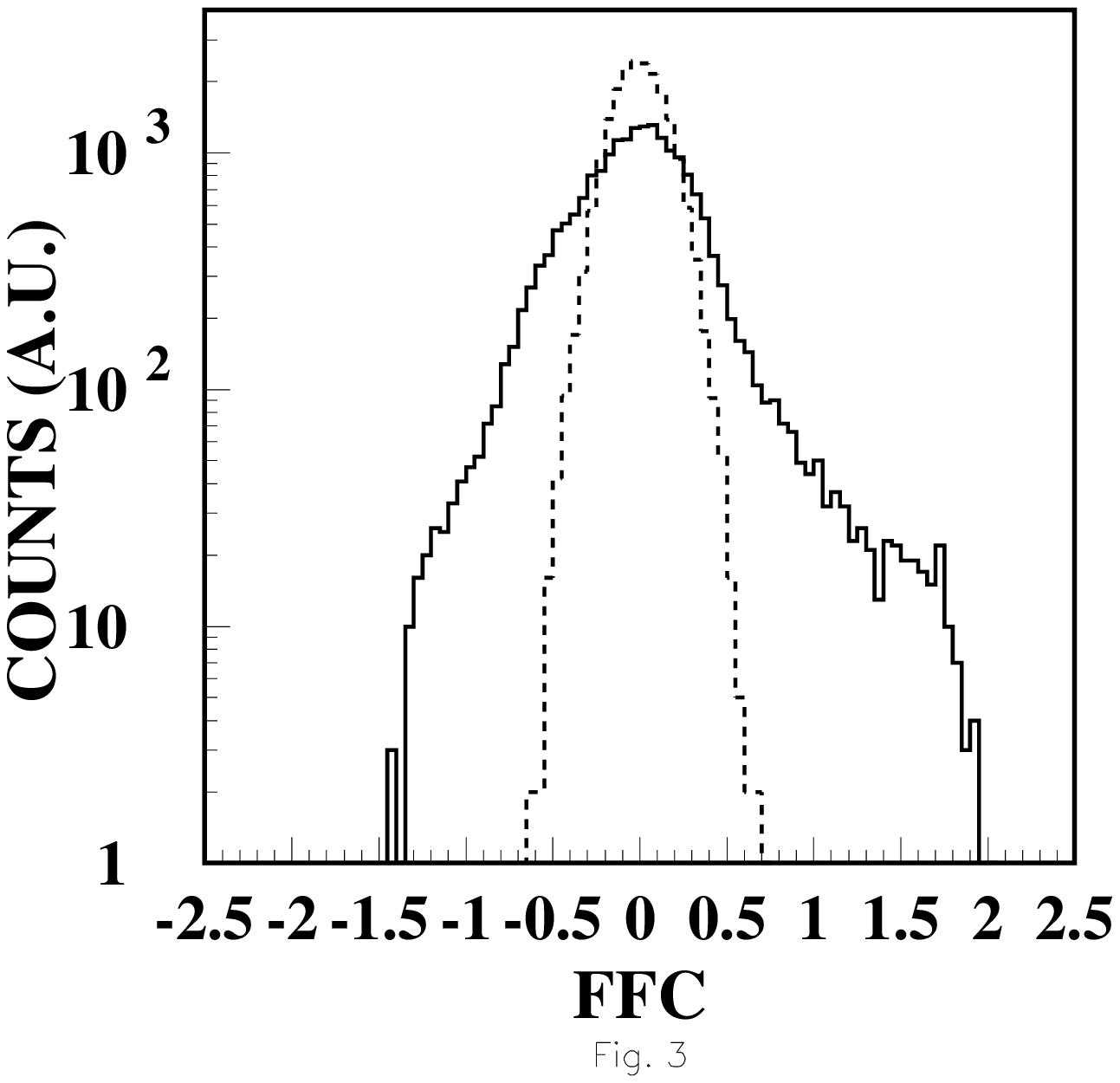}}
\newpage

\noindent ${}$\hbox{\epsffile{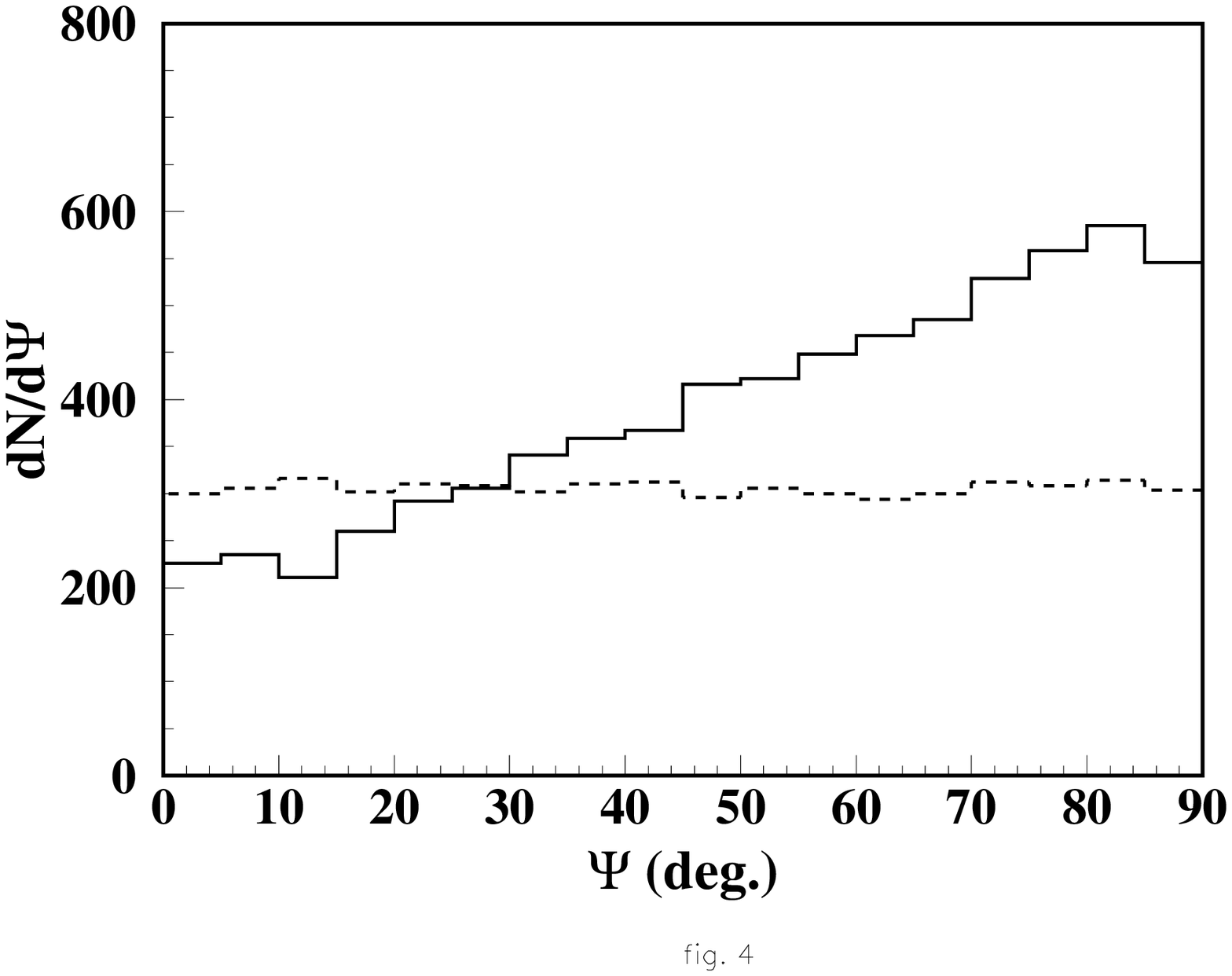}}



\end{document}